\pdfoutput=1

\documentclass[acmsmall]{acmart} 
\renewcommand\footnotetextcopyrightpermission[1]{} 
\AtBeginDocument{%
  }
    
\usepackage{algorithm}
\usepackage{algorithmic}

\usepackage{utfsym}

\usepackage{tabularx} 
\usepackage{array}

\usepackage{multirow} 
\usepackage{booktabs} 
\usepackage{makecell} 
\usepackage{array} 
\usepackage{multirow} 

\usepackage{ragged2e} 

\usepackage{caption}
\usepackage{array} 
\usepackage{cellspace} 
\usepackage{array} 
\usepackage{cellspace} 
\usepackage{colortbl} 
\definecolor{lightgray}{gray}{0.8} 

\usepackage[utf8]{inputenc}
\usepackage{listings}
\usepackage{xcolor}

\lstdefinelanguage{Solidity}{
    keywords={contract, function, public, payable, mapping, address, uint256, msg, sender}, 
    keywordstyle=\color{cyan!80!black}\bfseries, 
    morestring=[b]", 
    stringstyle=\color{orange!80!black}, 
    morecomment=[l]{//}, 
    morecomment=[s]{/*}{*/}, 
    commentstyle=\color{gray!60}\itshape, 
    sensitive=true 
}

\begin{document}
\title{Defending the Peg: Real-Time Dynamic Protection and Anomaly Detection in DeFi Stablecoins}

\author{Hengxing Zeng}
\authornotemark[1] 
\affiliation{%
  \institution{Hainan University}
  \city{Haikou}
  \country{China}}
\email{1434780680@qq.com}

\author{Shipeng Ye}
\authornote{These authors contributed equally to this work.}
\affiliation{%
  \institution{Hainan University}
  \city{Haikou}
  \country{China}}
\email{yeshipeng35@gmail.com}

\author{Xiaoqi Li}
\affiliation{%
  \institution{Hainan University}
  \city{Haikou}
  \country{China}}
\email{csxqli@ieee.org}

\renewcommand{\shortauthors}{Ye and Li}

\begin{abstract}
With the rapid evolution of the Decentralized Finance (DeFi) ecosystem, stablecoins have emerged as a critical infrastructure bridging the cryptocurrency market with traditional financial paradigms. However, stablecoin systems rely heavily on smart contracts to execute automated operations. The immutable nature of these systems post-deployment means that the exploitation of security vulnerabilities can lead to irreversible, massive economic losses and potentially trigger systemic financial risks. Current research on stablecoin smart contract security faces challenges such as a lack of domain-specific targeting and the obsolescence of static defense models. To address this, this paper systematically analyzes common attack vectors in stablecoin environments and proposes a practical, real-time dynamic defense architecture. By analyzing 12 real-world security incidents, we elucidate the underlying mechanisms of high-risk patterns such as reentrancy attacks, oracle manipulation, and composite flash loan attacks. Concurrently, we construct a real-time anomaly detection model utilizing multi-dimensional on-chain temporal features and the Bi-LSTM algorithm. Experimental results demonstrate that this model achieves a classification accuracy of 96.61\%, with an average recall rate of 97.70\% for malicious attack samples, and a single inference latency ranging from 1.5 to 2.8 milliseconds.
\end{abstract}



\maketitle
\fancyfoot{}
\pagestyle{plain} 

\section{Introduction}
With the continuous evolution of blockchain technology, the novel financial paradigm represented by Decentralized Finance has rapidly emerged, dismantling the geographical and institutional barriers of traditional finance and providing intermediary-free financial services globally ~\cite{aloudat2025metaverse,li2026uscsa,wang2026libscan}. As a core infrastructure of the DeFi ecosystem, stablecoins maintain value stability by pegging to fiat currencies or other assets. They effectively resolve the extreme volatility inherent in crypto assets, playing a crucial role as a medium of exchange, a store of value, and a form of collateral.

Recent market data indicates that by 2025, the total market capitalization of global stablecoins exceeded \$260 billion, accounting for over 70\% of the daily trading volume in the cryptocurrency market and serving as the primary source of liquidity \cite{wen2025stablecoins}. Stablecoin systems rely heavily on smart contracts to automate operations such as token minting, burning, collateral management, and liquidation without manual intervention. However, the post-deployment immutability of smart contracts dictates that exploited vulnerabilities can cause irreversible and catastrophic economic losses \cite{liu2025blockchain}. Historically, smart contract vulnerabilities have constituted a primary risk factor in the DeFi ecosystem. For instance, the infamous 2016 DAO attack, caused by a reentrancy vulnerability, led to the Ethereum hard fork ~\cite{liu2025reentrancy,li2026scpatcher,mo2025password}. Subsequent incidents, including the 2017 Parity multisig wallet exploit \cite{hossain2025bridging}, the 2018 Beauty Chain (BEC) overflow bug \cite{semwal2025security}, and the \$50 million exploit on the Binance Smart Chain (BSC) during a contract migration in 2021 \cite{chen2025security}, profoundly underscore the critical urgency of stablecoin smart contract security\cite{long2025fomo3d}.

Regarding specific vulnerability taxonomies, stablecoin smart contracts frequently encounter security flaws such as reentrancy attacks, integer overflows, access control defects, and oracle manipulation. Reentrancy attacks exploit external call mechanisms to repeatedly execute critical logic ~\cite{liu2025reentrancy,li2026psr2,liu2026liquilm}. Integer overflows cause anomalous asset calculations and illegal token issuance \cite{li2025understanding}. Access control defects grant attackers unauthorized privileges to hijack core system functions \cite{akon2025control}. Furthermore, oracle manipulation alters external price feeds to indirectly influence system behavior, serving as a primary risk for crypto-collateralized stablecoins. More alarmingly, these vulnerabilities are increasingly combined with novel financial instruments, such as flash loans, to orchestrate composite attacks \cite{wang2025prevention}. In a single, atomic, and risk-free transaction, attackers can complete capital acquisition, market manipulation, arbitrage, and loan repayment \cite{hu2026blockchain}. This atomic execution significantly amplifies both the efficiency and stealth of the attacks, posing profound challenges to conventional security paradigms\cite{wang2026ragas,gao2025implementation}.

Moreover, different categories of stablecoins exhibit distinct technical implementations and risk profiles. Fiat-collateralized stablecoins involve centralized operational management, presenting risks of administrative privilege abuse and access control deficiencies \cite{yan2025necessity}. Crypto-collateralized stablecoins depend heavily on oracles and liquidation mechanisms, making oracle manipulation their primary attack vector \cite{mahrous2025stablecoins}. Algorithmic stablecoins rely on supply-demand adjustments, facing severe risks of market manipulation and mechanism failures \cite{you2025hybrid}. The catastrophic collapse of Terra Luna in 2022 demonstrated how algorithmic design flaws, exacerbated by market panic, could vaporize over \$40 billion in market value and trigger industry-wide systemic contagion.

Under the current threat landscape, isolated vulnerability detection or static contract auditing is inadequate against complex attack scenarios \cite{tan2025advanced}. As composite attacks leveraging flash loans and oracle manipulation grow increasingly sophisticated, integrating artificial intelligence for real-time anomaly detection and dynamic monitoring has become an imperative trajectory for enhancing system security ~\cite{moradi2026smart,sun2025data,shi2025system}. Traditional pre-deployment static audits are fundamentally blind to novel, post-deployment composite attacks \cite{lin2026safety}. Conversely, a dynamic defense model can identify anomalous behaviors in real-time during an ongoing attack sequence and execute immediate interceptions, effectively bridging the inherent gaps of static auditing \cite{zhang2025active}.

Therefore, a systematic investigation into the security vulnerabilities, attack patterns, and dynamic defense mechanisms of stablecoin smart contracts not only holds substantial theoretical value but also carries critical practical significance for safeguarding the stability of the DeFi ecosystem and mitigating systemic financial risks.

The main contributions of this paper are as follows:
\begin{itemize}
\item \textbf{Systematic Threat Modeling for Stablecoin Ecosystems:} We systematically dissect the distinct technical mechanisms and risk profiles of fiat-collateralized, crypto-collateralized, and algorithmic stablecoins.
\item \textbf{Full-Lifecycle Dynamic Defense Architecture:} Moving beyond traditional static auditing, we propose a multi-layered, dynamic defense architecture that spans the pre-deployment, mid-execution, and post-incident phases.
\item \textbf{Real-Time Anomaly Detection via Mempool Monitoring:} We introduce a novel node-level defense mechanism deploying a Bidirectional Long Short-Term Memory network directly at the mempool layer. Our model effectively extracts multi-dimensional on-chain temporal features to intercept malicious payloads before block inclusion.
\end{itemize}
\vspace{-2ex}
\section{Background}

\subsection{Smart Contracts}
Smart contracts are self-executing protocols whose execution primarily relies on blockchain virtual machines for state updates and transaction processing ~\cite{lemayian2025evmx,lv2026mvcx,peng2026thought}. They utilize function calls to alter on-chain states, and the execution process typically involves inputting transactions, executing functions, updating states, and triggering events ~\cite{fan2026web3agent,yang2025pulling,ye2026chaindelta}. This execution paradigm is deterministic and immutable, and all nodes execute the contract code according to identical rules to maintain consistent outcomes. However, this also results in the long-term exploitability of code vulnerabilities. Once a contract is deployed, any inherent vulnerabilities persist indefinitely, unless the contract supports upgradability, they cannot be patched \cite{alsunaidi2025leveraging}. Within this execution model, smart contracts can perform external calls, a feature that lays the foundation for inter-contract interactions\cite{li2025interaction}. This allows disparate contracts to invoke one another, thereby facilitating the development of complex financial applications. Nevertheless, this mechanism concurrently introduces security risks. The primary issue is that upon invoking an external contract, execution sequence errors may occur during state updates, a scenario that constitutes the fundamental cause of reentrancy attacks\cite{liu2025reentrancy}.

Taking the classic "The DAO" attack as an example, attackers constructed a malicious contract that triggered a fallback function during the target contract's transfer operation ~\cite{ren2025lookahead,zhou2025blockchain}. The original function was repeatedly invoked before the state was updated, allowing the attackers to extract funds multiple times ~\cite{zhang2025smartreco,li2026systematic}. This case illustrates a critical security flaw in the smart contract execution model that a sequential dependency exists between state updates and external calls. In the Ethereum Virtual Machine (EVM) execution model, an external call suspends the execution of the current function and redirects control to the external contract's code \cite{wang2026empirical}. Only after the external call returns does the current function resume execution. Consequently, this affords attackers the opportunity to insert malicious logic and repeatedly invoke the current function before its state update is finalized.

Security risks also emerge when smart contracts perform numerical computations. Early versions of Solidity lacked automatic overflow checks, meaning integer overflows or underflows could result in erroneous balance calculations ~\cite{chen2025numscout,zhang2025novel}. For instance, during the minting or burning of tokens, if numerical ranges are not strictly bounded, attackers can construct extreme inputs to subvert system constraints. Specifically, when a larger number is subtracted from a smaller one, an integer underflow occurs, yielding a substantially large positive value \cite{akathoott2026sleek}. This enables attackers to bypass balance verification mechanisms and execute illicit fund withdrawals. Although the compiler has incorporated default overflow checks since Solidity version 0.8.0 \cite{tong2025sbugchecker}, such vulnerabilities remain prevalent in legacy contracts or in instances where developers have manually disabled these checks.

Access control mechanisms constitute another major security domain. Smart contracts typically employ modifiers to restrict permissions for sensitive operations, such as minting, burning, and contract upgrading, to ensure that only authorized administrators can invoke them ~\cite{liu2025have,zhang2025acf,baralla2025integrating}. However, if the permission architecture is flawed or lacks rigorous validation, attackers can invoke critical functions to seize control of the entire system \cite{liu2026dark}. Notably, the initialization functions of many contracts lack adequate permission checks \cite{li2025penetrating}. In such scenarios, attackers can preemptively execute the initialization function post-deployment, designate themselves as administrators, and subsequently take over the contract.
\vspace{-2ex}
\subsection{Stablecoin}

Based on their underlying value-backing mechanisms, stablecoins are primarily classified into three categories, including fiat-collateralized, crypto-collateralized, and algorithmic stablecoins, as in Figure~\ref{fig: classification}. These distinct types exhibit significant variations regarding smart contract design complexity and security risks, consequently presenting entirely different attack surfaces.
\vspace{-2ex}
\begin{figure}[h]
    \centering
    \includegraphics[width=0.65\textwidth]{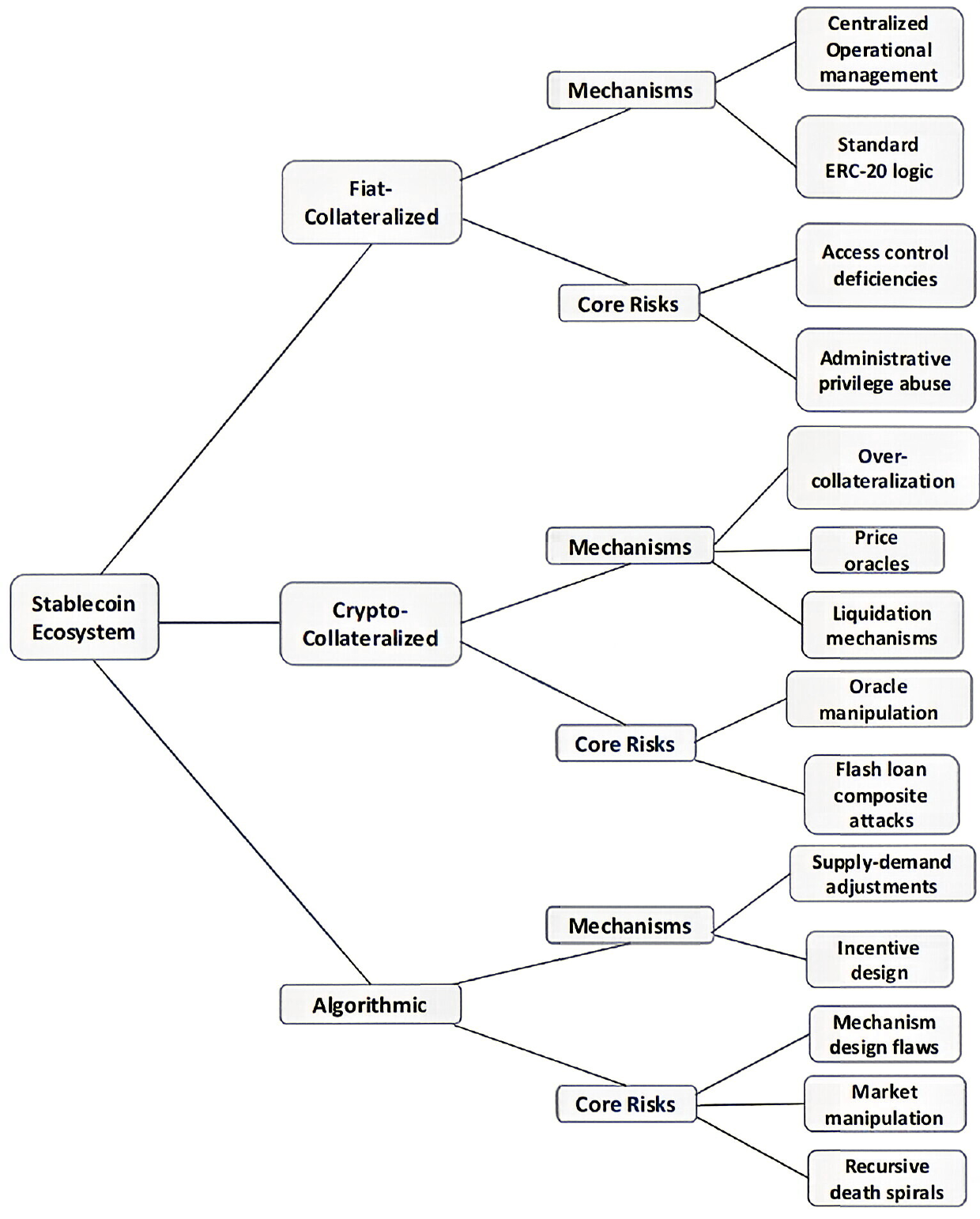}
    \caption{Taxonomy and Risk Profiles of the Stablecoin Ecosystem.}
    \vspace{-4ex}
    \label{fig: classification}
\end{figure}

\subsubsection{\textbf{Fiat-Collateralized Stablecoins}}
Fiat-collateralized stablecoins rely on centralized institutions to custody fiat currency and issue an equivalent volume of tokens on the blockchain. Users deposit fiat currency into the issuing institution's bank account, and the issuer subsequently mints an equivalent amount of stablecoins on-chain for the user \cite{lin2025parallel}. During a redemption operation, the issuer burns the on-chain stablecoins and transfers the fiat currency back to the user's bank account, thereby maintaining the price peg \cite{li2026sok}. The smart contract functionality for this category of stablecoins is relatively straightforward, primarily executing token minting, burning, and balance management. The core architecture typically adheres to the standard ERC-20 logic, augmented with basic access control mechanisms \cite{pathak2025performance}. However, the system's security risks are heavily concentrated in these access controls. Because the permissions for minting and burning are exclusively governed by a centralized issuer, any anomaly in permission management can result in severe security compromises. 

In several practical instances, flawed permission management designs have allowed attackers or malicious internal accounts to invoke critical administrative functions, resulting in illicit token issuance or parameter tampering \cite{tiwari2025mitigating}. Such vulnerabilities typically stem from deficient permission validation logic or ambiguous role-based access control (RBAC). For example, if the minting function fails to impose rigorous identity restrictions on the caller, attackers can effortlessly generate massive quantities of stablecoins, severely destabilizing the system \cite{li2025blockchain}. Furthermore, upgradable contract mechanisms introduce supplementary risks. In the event of an access control failure, attackers can exploit logic contract substitutions to assume total governance over the system. 

Consequently, the principal risks associated with fiat-collateralized stablecoins can be summarized as access control deficiencies and the abuse of centralized privileges.

\subsubsection{\textbf{Crypto-Collateralized Stablecoins}}

Crypto-collateralized stablecoins, epitomized by DAI, operate on the core logic of generating stablecoins by collateralizing mainstream cryptocurrency assets. This model typically necessitates over-collateralization to mitigate the high volatility of crypto asset prices \cite{yan2025necessity}. For instance, a user might collateralize \$150 worth of ETH to generate \$100 in stablecoins, with the excess collateral serving as a buffer against ETH price depreciation. The core mechanisms encompass collateral management, loan generation, collateralization ratio computation, and liquidation protocols. When the price of the collateralized asset drops—causing the collateralization ratio to fall below a predefined safety threshold—the system automatically liquidates the user's collateral to repay the debt, thereby ensuring systemic solvency.

In this paradigm, smart contracts manage complex state transitions and computations, relying heavily on price oracles to supply real-time asset price data. Because the system requires instantaneous collateral prices to calculate ratios and determine liquidation triggers, this external dependency constitutes a primary attack vector. Since oracle price data is sourced from off-chain entities or cross-chain decentralized exchanges (DEXs), attackers can manipulate these data feeds to subvert the system's evaluative logic \cite{moghadam2026signet}. Several paradigmatic attack cases demonstrate that malicious actors can manipulate asset prices on DEXs, distort oracle data, and ultimately alter collateralization ratio computations. In certain exploits, attackers execute large-volume trades to artificially inflate an asset's price within a narrow time window. This deceives the system into miscalculating the asset's collateral value as artificially high, enabling the attacker to borrow stablecoins far exceeding legitimate collateral constraints before swiftly exiting the protocol.

Attackers frequently leverage flash loans to close the loop on these attack vectors \cite{huang2026flashshield}. First, they utilize a flash loan to acquire massive liquidity, allowing them to borrow hundreds of millions of dollars in a single transaction without upfront collateral. Second, they manipulate prices in low-liquidity markets by deploying these borrowed funds to rapidly purchase assets and artificially drive up prices. Third, this price manipulation cascades into the oracle data, causing the system to perceive the collateral's price as extraordinarily high. Finally, this triggers the borrowing or liquidation logic, allowing the attacker to borrow a massive volume of stablecoins against the overvalued collateral and repay the flash loan within the same transaction, thereby realizing a substantial arbitrage profit.

\subsubsection{\textbf{Algorithmic Stablecoins}}

Algorithmic stablecoins operate without relying on collateralized assets. Instead, they regulate token supply via smart contracts to achieve price stability. Their core mechanisms include price feedback systems, inflation and deflation adjustments, and incentive design. When the stablecoin's price exceeds its peg, the system mints additional tokens, diluting the supply to drive the price down. Conversely, when the price falls below the peg, the system burns tokens, restricting supply to drive the price up. The system theoretically maintains price stability through this algorithmic supply-and-demand dynamic.

However, these stablecoins face substantial operational risks. In a typical failure scenario, a decline in market confidence triggers user sell-offs, causing the price to drop. The system attempts to maintain the peg by minting more tokens, but this hyperinflationary issuance further dilutes value, exacerbating the price decline and intensifying panic selling. This creates a recursive downward spiral that ultimately leads to systemic collapse. A highly representative incident is the 2022 Terra (LUNA) crash, wherein the price of UST plummeted from its peg to less than \$0.01 within a matter of days. This resulted in a total systemic collapse, inflicting losses exceeding \$40 billion. Analyzed from a technical perspective, such failures are not merely the result of codebase vulnerabilities, but rather the consequence of the tight coupling between economic model design and contract execution logic. The primary catalyst for the Terra collapse was the inherent structural deficiency of the uncollateralized algorithmic stablecoin model, compounded by panic selling triggered by a total collapse in market confidence. Because algorithmic stablecoins rely on highly complex contract logic, vulnerabilities can frequently exist at the code level, which further amplifies the overall risk profile. Therefore, the core risk of algorithmic stablecoins lies in the interaction between mechanism design flaws and the amplifying effects of market behavior. This category of stablecoins possesses the broadest attack surface, encompassing both code-level vulnerabilities and economic mechanism exploits.

\subsection{Smart Contract Security Threat Model}

\subsubsection{\textbf{Attack Surface Analysis}}

Based on practical cases, the attack surfaces of stablecoin systems can be categorized into the following domains:

\begin{itemize}

\item \textbf{Code-Level Vulnerabilities}: This includes reentrancy attacks and integer overflows, which are directly exploited due to flaws inherent in the contract's execution logic.

\item \textbf{Access Control Vulnerabilities}: Instances such as unauthorized access allow attackers to govern critical system functions. These vulnerabilities are predominantly found in fiat-collateralized stablecoins, where attackers can directly hijack core administrative functions.

\item \textbf{External Dependency Risks}: A prime example is oracle manipulation, where attackers indirectly compromise the entire system by maliciously influencing input data. Such vulnerabilities frequently manifest in crypto-collateralized stablecoins due to the system's heavy reliance on external price feeds.

\item \textbf{Economic Model Vulnerabilities}: This involves failures in algorithmic stability driven by incentive mechanisms or market behaviors. These vulnerabilities predominantly exist in algorithmic stablecoins, where attackers exploit structural defects within the economic model itself.

\item \textbf{Composite Attack Vectors}: These involve the synergistic application of multiple techniques, such as the combination of flash loans, oracle manipulation, and liquidation attacks. Attackers chain multiple vulnerabilities together to achieve sophisticated and devastating attack outcomes.
    
\end{itemize}

\subsubsection{\textbf{Typical Attack Paths}}

\begin{figure}[h]
    \centering
    \includegraphics[width=0.65\textwidth]{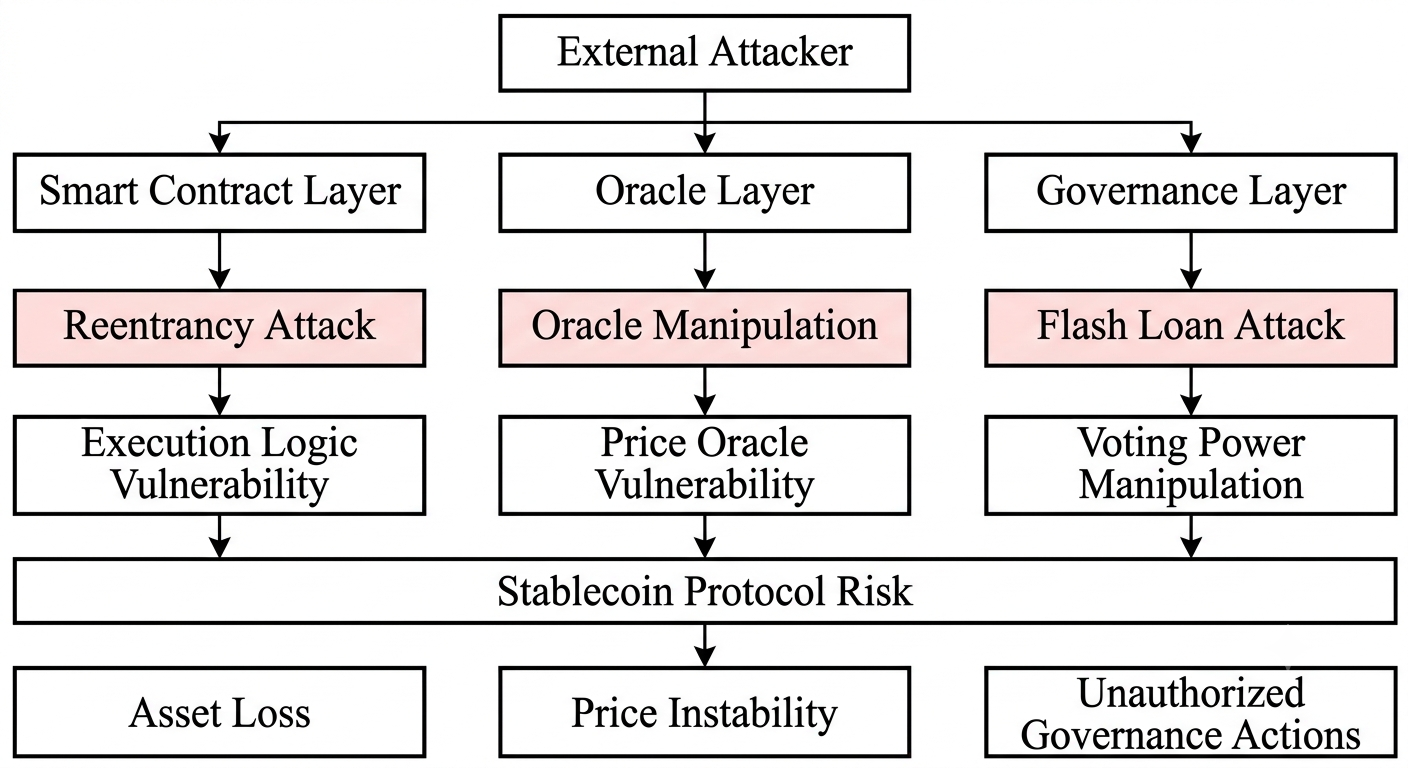}
    \caption{Threat Model of Security Attacks in Stablecoin Systems.}
    \label{fig: thread_model}
\end{figure}

Based on empirical cases, typical attack paths within stablecoin systems can be abstracted, as shown in Figure ~\ref{fig: thread_model}. These trajectories recursively appear across numerous attack incidents and exhibit significant commonalities:

\begin{itemize}

\item \textbf{Reentrancy Attack Path}: The execution sequence typically initiates with an external call, followed by the activation of a fallback function \cite{wei2025veriexploit}. This enables a recursive invocation of the target function prior to state updates, ultimately culminating in unauthorized fund extraction.

\item \textbf{Oracle Manipulation Path}: This exploit generally begins with targeted market manipulation that induces a significant asset price deviation \cite{gao2025airaclex}. The distorted data is subsequently ingested during an oracle update, leading to a critical system miscalculation.

\item \textbf{Flash Loan Composite Attack Path}: This trajectory commences with massive fund acquisition via a flash loan without upfront collateral \cite{zhang2025following}. The acquired liquidity is then deployed for market manipulation to trigger vulnerable contract logic, concluding with a profitable exit and loan repayment within the same atomic transaction.

\item \textbf{Privilege Abuse Path}: This path involves initial unauthorized privilege acquisition, which facilitates subsequent arbitrary parameter modification, thereby granting the attacker complete control over the system's behavior \cite{jarkas2025container}.
    
\end{itemize}

\subsection{Security Protection Technologies}

Addressing the aforementioned attack patterns, existing security protection technologies are primarily distributed across various phases of the development lifecycle. However, empirical attack cases have revealed their inherent limitations, demonstrating an inability to fully counter the sophisticated exploits prevalent in stablecoin scenarios.

\subsubsection{\textbf{Static Analysis Methods}}

Static analysis is a vulnerability detection method targeting contract source code or bytecode \cite{gomes2025static}. It detects vulnerable behavioral patterns by analyzing the code's control flow and data flow, thereby uncovering predefined vulnerabilities. The primary advantage of static analysis lies in its ease of automation and its efficacy in identifying vulnerabilities within a given set of contract code. Nevertheless, static analysis possesses distinct limitations. First, it cannot detect vulnerabilities that emerge dynamically during contract execution. For instance, when a contract utilizes external data for computations, malicious data inputs might trigger vulnerabilities that static analysis typically fails to capture. Second, static analysis is often ineffective in evaluating dynamic calls or complex logic within the contract.

\subsubsection{\textbf{Fuzzing Methods}}

Fuzzing conducts vulnerability detection by injecting random, invalid, or malicious data into a contract \cite{kong2025smart}. This method is capable of uncovering edge cases within smart contracts and identifying potential security issues. However, in practical applications, neither dynamic analysis nor fuzzing can achieve complete code path coverage. In complex contracts, certain vulnerabilities are inevitably overlooked, and the results often suffer from false positives and false negatives, thereby complicating the analysis process.

\subsubsection{\textbf{Formal Verification Methods}}

Formal verification for smart contract vulnerabilities is a technique rooted in mathematics and logical reasoning, designed to mathematically guarantee the absence of security flaws during contract execution \cite{huang2025enhancing}. Smart contracts can employ formal verification across their lifecycle, utilizing this technique throughout both the design and development phases. Code-level formal verification can implement dual verification within a unified environment using both source code and compiled bytecode. The source code undergoes transformation for verification, while the compiled bytecode is decompiled to verify low-level properties. These two verification approaches utilize equivalence proofs to guarantee functional and operational consistency.

Although this method provides robust security guarantees, its modeling and proving processes are highly complex, leading to increased difficulties in implementation and comprehension. Consequently, its application in stablecoin systems is limited. Formal verification can only prove properties at the code level, it cannot mathematically guarantee the security of the economic model, thus rendering it incapable of preventing economic-layer exploits such as oracle manipulation or composite flash loan attacks.

\subsubsection{\textbf{Intermediate Representation}}

Intermediate Representation (IR) is a data structure utilized to represent a program's intermediate state or form \cite{chen2025arkanalyzer}. It is typically generated after the source code undergoes lexical and syntax analysis phases, and is employed prior to target code generation, remaining independent of specific programming languages and target platforms. Common intermediate representations include Abstract Syntax Trees (AST), Three-Address Code, and Static Single Assignment (SSA) form. Different compilers and toolchains may employ distinct IRs. Within a compiler, the IR plays a crucial role during the optimization and analysis phases. By analyzing and transforming the IR, various optimization measures, such as constant folding, loop optimization, and inlining, can be implemented. Simultaneously, IR can also be utilized to generate target code, converting the optimized program into platform-specific opcodes or bytecode. Smart contracts are generally written in high-level languages. However, due to the complex structure and semantics of the source code, direct vulnerability detection on the source code often encounters significant challenges. Therefore, translating smart contract source code into an intermediate representation can simplify the vulnerability detection process and facilitate the application of static analysis techniques.

\subsubsection{\textbf{Machine Learning Methods}}

The objective of machine learning is to enable computers to make predictions and decisions to solve problems based on past experience, through data learning and pattern recognition, without requiring explicit programming instructions. In recent years, machine learning has been increasingly adopted for smart contract vulnerability detection, capable of automatically mining potential vulnerabilities and security issues by learning from vast datasets of smart contracts \cite{alsunaidi2025leveraging}. During the training process, algorithms automatically discover patterns and regularities within the data, using them as a basis to adjust and update model parameters. Through continuous iterative learning, the model can be progressively optimized and improved, enhancing its generalization capability on unseen data. Machine learning plays a crucial role in various aspects of smart contract vulnerability detection, including feature extraction, anomaly detection, vulnerability classification, and generative training. At the same time, applying machine learning to this task still faces significant challenges. The complexity and highly dynamic nature of smart contracts make vulnerability detection increasingly difficult. Furthermore, the lack of sufficient training data and high-quality labeled data also constrains the efficacy of model training.

\section{Defense Methods for Stablecoin Smart Contracts}

\subsection{Full-Lifecycle Dynamic Defense Architecture}

Stablecoin smart contracts face complex security threats, and traditional static defense methods generally fail to keep pace with real-time on-chain attacks \cite{mishra2025blockchain}. To address this, this paper proposes a defense architecture, as shown in Figure~\ref{fig: classification} comprising pre-deployment screening, mid-execution interception, and post-incident response. This model leverages the synergy of multiple technologies to integrate security defenses throughout the complete lifecycle of smart contracts.

\begin{figure}[h]
    \centering
    \includegraphics[width=0.8\textwidth]{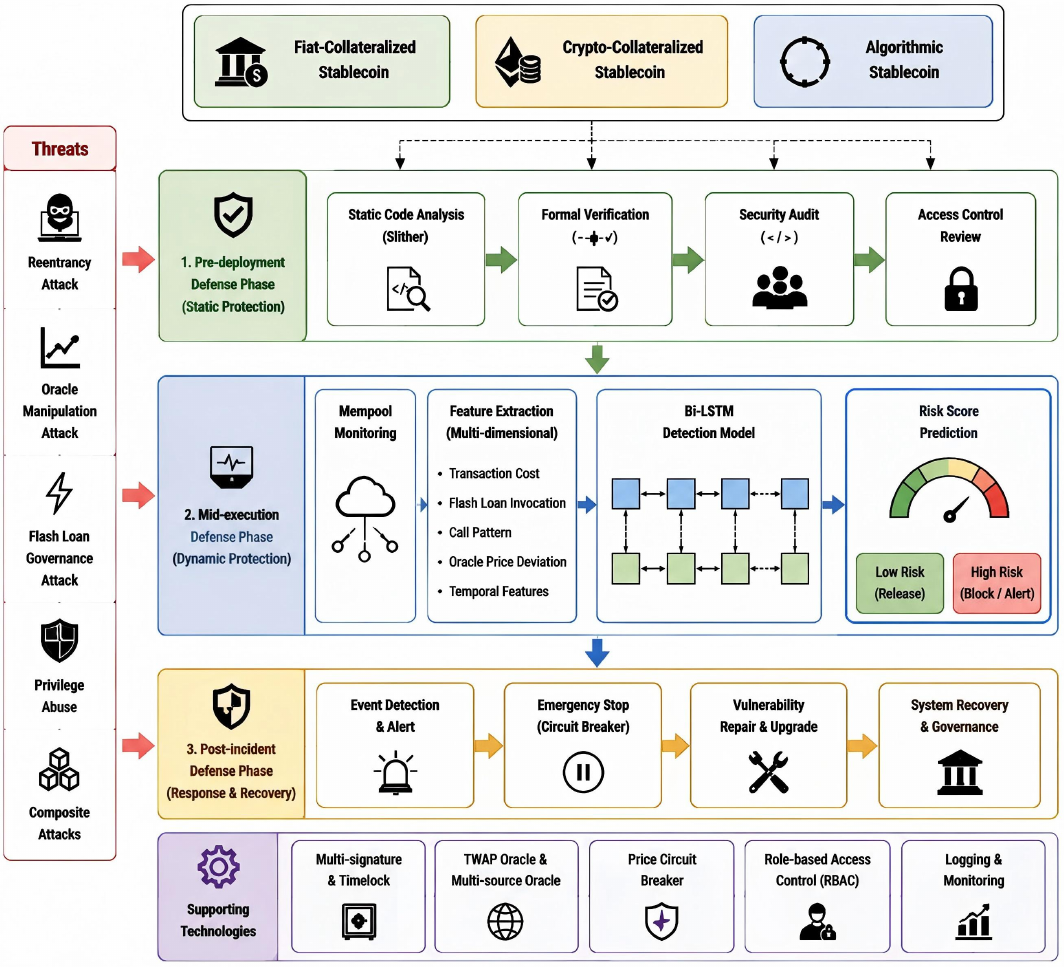}
    \caption{Overview of the proposed lifecycle-based stablecoin security framework.}
    \label{fig: classification}
\end{figure}

\subsubsection{\textbf{Pre-Deployment Defense Phase}}

Before the contract code is deployed to the mainnet, this phase primarily focuses on eliminating existing defects in code logic and business semantics. This stage utilizes static code scanning to detect common vulnerabilities such as integer overflows and reentrancy risks, and applies formal verification techniques to mathematically prove that the core business logic maintains state consistency. Concurrently, automated and manual security audits conducted by professional third-party organizations establish a fundamental security baseline prior to protocol launch, thereby mitigating potential risks at the underlying code level.

\subsubsection{\textbf{Mid-Execution Defense Phase}}

During the deployment and operational phases of the contract, the focus shifts to addressing zero-day vulnerabilities and complex composite attacks. This phase designates the node memory pool (Mempool) as the central monitoring layer and employs deep learning methods, such as Bi-LSTM, to construct an anomaly detection engine. This engine performs millisecond-level temporal feature extraction and risk inference on transaction sequences that have not yet been packed into the blockchain. Upon detecting attack payloads with a high probability of risk, the system automatically triggers an RPC-level hard interception or issues a suspend command to the target contract. This prevents the attack from causing substantial state transitions and successfully blocks the threat.

\subsubsection{\textbf{Post-Incident Defense Phase}}

In the event of a security breach under extreme circumstances, the priority is to contain the scope of losses and facilitate business recovery. This phase primarily relies on global pause (circuit breaker) functionalities and emergency proxy upgrade mechanisms predefined during the initial contract design. Utilizing an event-driven on-chain state monitoring model, if an unexpected massive outflow of funds occurs, the system automatically activates contingency plans to isolate the compromised liquidity pool. Subsequently, by employing automated vulnerability repair assistive technologies combined with manual intervention, the vulnerabilities can be rapidly patched, restoring the system to normal operational status.

\subsection{Differential Defense Strategies}

Given the distinct risk profiles and attack surfaces inherent in various stablecoin architectures, this paper proposes differential defense strategies tailored to the three primary stablecoin paradigms, systematically mitigating their respective core vulnerabilities.

\subsubsection{\textbf{Privilege Isolation and Auditing}}

The primary risks associated with fiat-collateralized stablecoins are access control deficiencies and privilege abuse; thus, defense efforts must center on rigorous permission management.

\begin{itemize}

\item \textbf{Granular RBAC}: Decouple permissions for minting, burning, and upgrading into distinct roles to eliminate the existence of a super-administrator, ensuring each role adheres to the principle of least privilege.

\item \textbf{Multi-Signature and Time-Lock Mechanisms}: Mandate multi-signature approvals for all sensitive operations and integrate time locks to enforce a 24 to 48 hour execution delay. This provides users with a sufficient withdrawal window, effectively deterring the immediate execution of malicious or compromised administrative commands.

\item \textbf{Periodic Permission Audits}: Conduct routine inspections of contract permission configurations to verify the absence of unauthorized functions, thereby preventing privilege escalation vulnerabilities.

\item \textbf{Transparent Reserve Assets}: Require the regular cryptographic disclosure of proof of reserves to prevent the misappropriation of collateralized funds by the issuing entity.
    
\end{itemize}

\subsubsection{\textbf{Oracle Security and Liquidation Protection}}

The core vulnerability of crypto-collateralized stablecoins lies in oracle manipulation. Thus, defense mechanisms must prioritize oracle data integrity.

\begin{itemize}

\item \textbf{Mandatory TWAP (Time-Weighted Average Price)}: Enforce the use of TWAP and strictly prohibit reliance on spot prices, fundamentally mitigating the system's susceptibility to flash price manipulation.

\item \textbf{Multi-Source Oracle Aggregation}: Deploy mature decentralized oracles to aggregate price feeds from multiple independent data sources, thereby eliminating the single-point-of-failure risks associated with centralized data feeds.

\item \textbf{Price Circuit Breakers}: Automatically suspend liquidation and borrowing functionalities when asset price volatility exceeds a predefined threshold, defending the protocol against extreme price manipulation events.

\item \textbf{Liquidation Safeguards}: Impose strict upper bounds on the maximum volume of a single liquidation event to prevent attackers from exploiting liquidation mechanics for predatory arbitrage.
    
\end{itemize}

\subsubsection{\textbf{Economic Model Stress Testing and Mechanism Defenses}}

The critical risks characterizing algorithmic stablecoins are inherent economic design flaws and the propensity for recursive death spirals. Therefore, protective strategies must focus on the resilience of the economic model.

\begin{itemize}

\item \textbf{Economic Model Stress Testing}: Prior to deployment, subject the economic model to rigorous computational stress testing under extreme scenarios to simulate black swan market events. This evaluates model stability and ensures resilience against death spirals during severe market distress.

\item \textbf{Algorithmic Circuit Breakers}: Automatically halt token supply adjustments when the stablecoin's price deviates from its peg beyond a critical threshold, thereby arresting the systemic feedback loop that exacerbates a death spiral.

\item \textbf{Code-Level Security Auditing}: Given the high structural complexity of algorithmic stablecoin logic, mandate exhaustive and rigorous code audits to eradicate underlying vulnerabilities such as reentrancy and integer overflows.

\item \textbf{Incentive Mechanism Optimization}: Refine the incentive structures governing algorithmic supply-and-demand regulation to preclude attackers from exploiting these mechanisms for malicious, risk-free arbitrage.
    
\end{itemize}

\subsection{AI-Based Real-Time Anomaly Detection System}

Flash loan composite attacks and oracle manipulation events, which frequently occur within the smart contract ecosystem, exhibit formidable destructive potential. Under the operational paradigms of mainstream blockchains such as Ethereum, conventional defense mechanisms primarily rely on retrospective smart contract security audits or blacklist-based interception using the historical invocation records of known malicious addresses. However, contemporary composite attacks are typically executed within an extremely narrow time window, seamlessly completing the entire lifecycle, from initial fundraising and multi-path cross-contract invocations to the manipulation of underlying liquidity pool prices and the final risk-free profitable exit. This instantaneous nature renders transition responses largely ineffective.

\begin{figure}[h]
    \centering
    \includegraphics[width=0.8\textwidth]{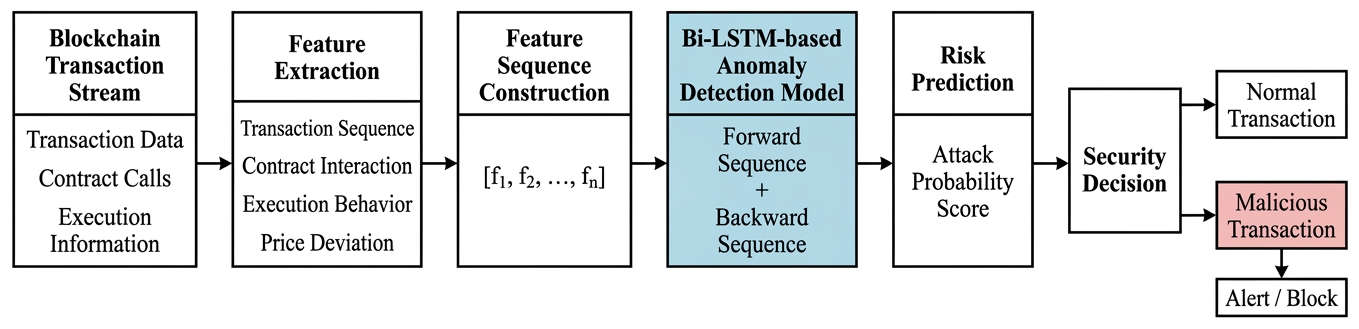}
    \caption{AI-based Real-time Transaction Detection Workflow for Stablecoin Security.}
    \label{fig: AI_Bi_LSTM}
\end{figure}

In this paper, we design a preemptive AI anomaly detection system deployed at the Mempool level of underlying blockchain nodes. To process pending transactions prior to the execution of state transitions by the Ethereum Virtual Machine (EVM), this system utilizes a deep neural network to extract temporal features from invocation sequences, as shown in Figure ~\ref{fig: AI_Bi_LSTM}. It performs millisecond-level preemptive inference to facilitate precise, physical-level blocking of malicious transactions before they are recorded on the chain.

\subsubsection{\textbf{Multidimensional Reconstruction of On-Chain Interaction Feature Engineering}}

Malicious on-chain transactions often manifest as fund fragmentation and redundant nested transfers, utilizing these patterns to evade static rule-based detection methods. In this paper, we construct a transaction feature vector space encompassing five dimensions to capture the primary characteristics of attack behaviors. Utilizing statistical principles, this study generates a simulated on-chain dataset comprising 100,000 samples for model training and validation, with a positive-to-negative sample ratio set at 95\% normal transactions to 5\% malicious attacks. These five-dimensional features are categorized into three major classes based on business logic. The specific extraction contents and selection criteria are detailed as follows:

\begin{itemize}

\item \textbf{Basic Transaction and Execution Cost Features}: The extracted features include the transaction principal size (Amount), the maximum gas fee consumed (Gas Limit), and the encoded length of the input data payload (Input Data Length). The rationale for selecting these features is that flash loan composite attacks necessitate cross-contract invocations across multiple DeFi protocols, which typically involve deep function nesting, external state calls, and multiple modifications via the underlying storage instruction (SSTORE). Compared to routine on-chain transfers or single DEX token swaps, attack transactions consume significantly more gas and feature longer input code payloads. To secure substantial arbitrage profits, the principal volume transferred is generally massive. Attackers may attempt to obfuscate these three metrics to masquerade as benign transactions. In the experiments, these three features were modeled using a normal distribution with a large variance, a mean of 0.5, and a standard deviation of 0.3. This distribution ensures an overlap between normal and attack samples across basic metrics, dictating that the model cannot rely solely on a single numerical threshold for judgment, and must integrate temporal context for accurate classification.

\item \textbf{Liquidity Flash Loan Invocation Features}: This extracted feature determines whether a flash loan interface is triggered within the transaction and its internal call stack. The selection is based on the premise that oracle manipulation relies heavily on massive, uncollateralized capital to disrupt market price equilibrium. Flash loans grant attackers access to substantial liquidity at a minimal cost, serving as the primary funding source for most serial composite attacks. Thus, this paper identifies it as a critical dimension. In reality, legitimate high-frequency quantitative arbitrage bots also legally invoke flash loans to eliminate price discrepancies across DEXs. Therefore, this feature is modeled using a binomial distribution probability. It is assumed that attackers, requiring high leverage, have an 85\% probability of utilizing flash loans, whereas normal arbitrageurs possess an 8\% probability of legitimate invocation. This probability assignment objectively reflects the presence of confounding factors in real-world business logic and prevents the model from falling into the overfitting fallacy of equating any flash loan usage with an attack.

\item \textbf{Oracle Price Deviation Features}: The extracted features encompass the anticipatory calculation of the underlying liquidity pool's AMM (Automated Market Maker) token swap price slippage that the current transaction execution might induce. This selection is grounded in the constant product formula of AMMs, where a massive unilateral selling pressure within a short timeframe distorts the exchange ratio. A significant price deviation is the primary consequence of an attacker executing a "buy low, sell high" strategy to exploit oracle price discrepancies, serving as a vital quantitative indicator of malicious manipulation. Normal transactions generate relatively minor price slippage. The experiment simulates this using an exponential distribution with a scale parameter set to 0.15, indicating that while most arbitrage slippages are minute, extreme deviations occur under a long-tail effect. To guarantee an arbitrage margin, malicious attacks induce pronounced price anomalies. This paper analyzes this using a normal distribution model. Extracting this feature quantifies the extent of damage the transaction inflicts on market prices.
    
\end{itemize}

\subsubsection{\textbf{Cost-Sensitive Detection Model Based on Bi-LSTM}}

Addressing the aforementioned feature sequences characterized by temporal dependencies and distributional noise, traditional static rules or shallow machine learning algorithms struggle to effectively capture contextual relationships across the temporal dimension. In this paper, we adopt the Bidirectional Long Short-Term Memory (Bi-LSTM) network as the core architecture of the anomaly detection model.

Bi-LSTM comprises both forward and backward layers for information propagation. Within a complex composite attack, a complete transaction typically entails early-stage borrowing for fundraising, mid-stage price manipulation, and late-stage account reconciliation and withdrawal. The Bi-LSTM can concurrently extract the bidirectional temporal correlations of these sequential steps, thereby more accurately capturing the complete transactional semantics. The system configures the input dimension and establishes a 2-layer bidirectional hidden network, setting the hidden node dimension to 16. The network extracts the output of the sequence's final time step, processing it through a fully connected layer and a Sigmoid activation function to output a continuous numerical value ranging from 0 to 1, representing the probability of anomalous risk for the transaction.

On-chain data inherently exhibits a state of class imbalance between positive and negative samples. As previously established, normal transactions account for 95\% of the total dataset. If a conventional cross-entropy loss function is employed directly for network training, the model is prone to converging into a local optimum. Specifically, leaning towards predicting the vast majority of samples as normal to achieve a high overall Accuracy. This numerical bias renders the model highly inefficient at identifying sparse attack samples, causing the recall rate to approach zero, thus stripping the model of its practical defensive utility.

To address this imbalance at the algorithmic level, this paper introduces a Cost-Sensitive Learning paradigm. During the computation of backpropagation gradients, the system utilizes a binary cross-entropy loss function incorporating a positive sample penalty weight, calibrated according to the a priori ratio of normal to malicious samples. Within the defined experimental distribution, malicious samples are assigned a misclassification penalty weight approximately 19 times higher than that of normal samples. The learning rate of the Adam optimizer is set to 0.005, increasing the cost of misclassifying minority class samples during the training process. By adopting this strategy, this paper prioritizes enhancing the recall rate for anomalous transactions, even at the expense of a marginal reduction in precision, thereby minimizing the risk of false negatives in practical deployments.

\subsubsection{\textbf{Automated Interception Mechanism Based on Mempool Monitoring}}

During the inference phase, the neural network model predominantly performs matrix operations characterized by low computational latency, establishing the feasibility of deploying it on Ethereum Remote Procedure Call (RPC) nodes or validator nodes.

This paper deploys the anomaly detection engine at the ingress of the node's transaction queue (Tx-Queue). A real-time mempool monitoring daemon, monitor\_mempool, is designed to validate its operational efficacy. This daemon continuously simulates the influx of on-chain transaction data, dynamically injecting feature fluctuations that align with the previously established statistical distributions during runtime. To account for underlying inference latency, the system introduces a random temporal jitter ranging from 0.15 to 0.22 seconds, simulating data transmission and deserialization delays inherent in real-world distributed networks.

Once the node captures and extracts the transaction features in real time, the data is fed into the Bi-LSTM network loaded with pre-trained weights. Operating with gradient computation disabled (torch.no\_grad()), the model performs forward propagation and outputs the systemic risk confidence score for the transaction. To mitigate the false positive rate associated with legitimate high-frequency arbitrage operations, the system sets the risk interception threshold at 85\%. Based on this threshold, the system defines the following two automated processing protocols:

\begin{itemize}

\item \textbf{Normal Release}: When the attack confidence is below 85.00\%, the system determines that the current transaction conforms to standard on-chain interaction patterns. This spectrum includes ordinary address transfers and conventional quantitative arbitrage trades that may induce slight price slippage. In this state, the system merely records a release log in the background and broadcasts the transaction data normally to the peer-to-peer (P2P) network or forwards it to the miner node pool to await consensus packing, without interfering with its standard execution flow.

\item \textbf{High-Risk Alert and Automated Interception Closed-Loop}: When the attack confidence reaches or exceeds 85.00\%, the model detects anomalies such as irregular gas consumption, flash loan invocations, and oracle price deviations during its comprehensive analysis. Within a sub-second window, the system initiates an automated closed-loop response consisting of Local Mempool Filtering and On-Chain Automated Defense.
    
\end{itemize}

\section{Experiment Evaluation}

In this section, we evaluate the effectiveness of the proposed stablecoin security framework through two groups of experiments. The first experiment focuses on validating the effectiveness of the proposed smart contract defense mechanisms against representative stablecoin-related attacks. The second experiment evaluates the performance of the proposed AI-based real-time anomaly detection system in identifying malicious transaction behaviors. Specifically, we aim to answer the following research questions:

\textbf{RQ1:} Can the proposed defense mechanisms effectively mitigate representative attacks targeting stablecoin smart contracts?

\textbf{RQ2:} Can the proposed Bi-LSTM-based anomaly detection model accurately identify malicious transactions while maintaining low inference latency?

To answer these questions, we construct a local Ethereum-compatible experimental environment and conduct attack reproduction experiments, defense verification experiments, and AI model performance evaluations.

\subsection{Experimental Setup}

All experiments are conducted on a local Ethereum-compatible testing environment. The smart contract experiments are implemented using Solidity and executed on the Foundry framework. The AI-based anomaly detection model is developed using PyTorch. The detailed hardware and software configurations are summarized in Table~\ref{tab:environment}.

\begin{table}[h]
\centering
\caption{Experimental Environment Configuration}
\label{tab:environment}
\begin{tabular}{ll}
\hline
Component & Configuration \\
\hline
Operating System & Ubuntu 22.04 \\
Blockchain Framework & Foundry v1.5.1 \\
Smart Contract Language & Solidity \\
AI Framework & PyTorch \\
Python Version & 3.12.3 \\
CPU & 16-core Processor \\
Memory & 32 GB RAM \\
\hline
\end{tabular}
\end{table}

\begin{figure}[h]
    \centering
    \includegraphics[width=0.8\textwidth]{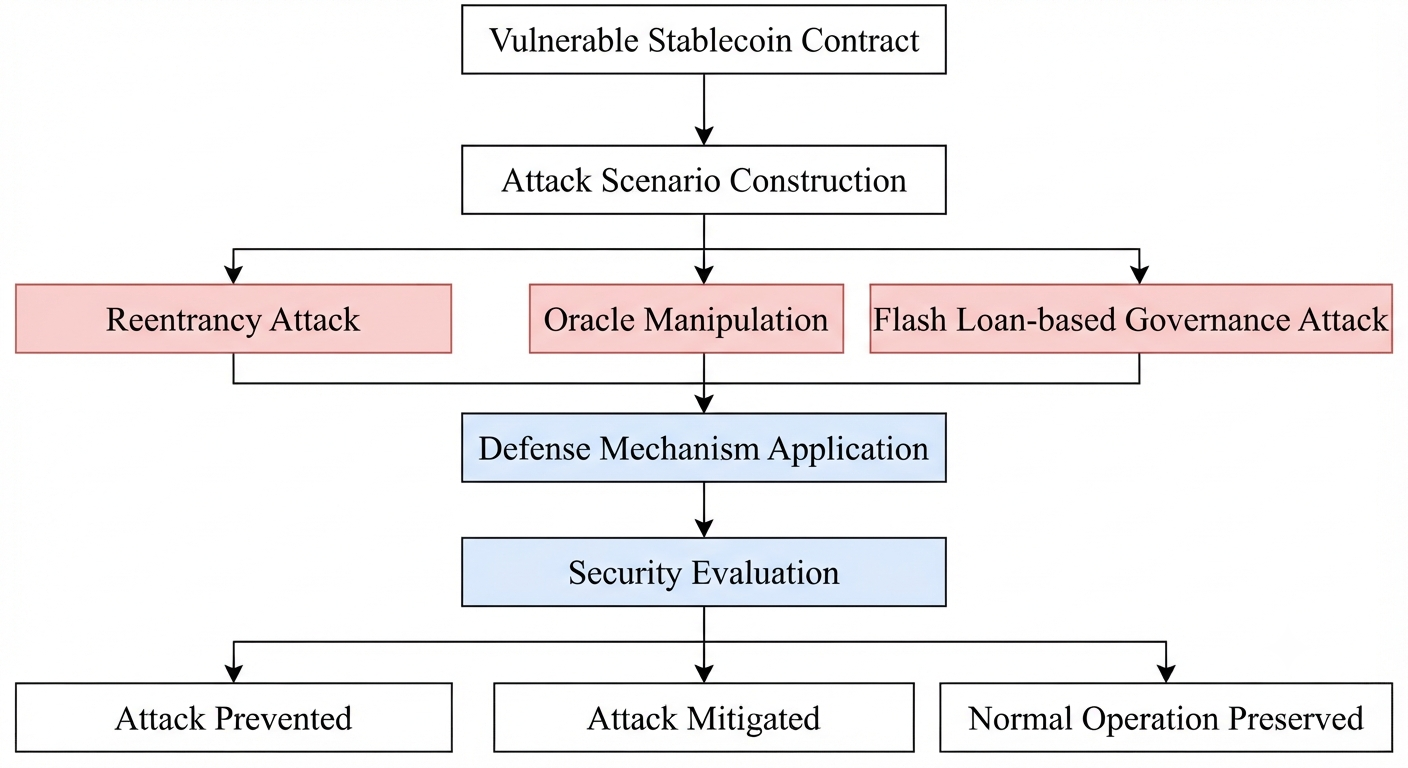}
    \caption{Experimental Workflow for Stablecoin Attack Defense Evaluation.}
    \label{fig: Workflow_for_Stablecoin}
\end{figure}

All smart contract experiments are conducted in a local Ethereum-compatible testing environment based on Foundry. Foundry is selected because it provides efficient smart contract compilation, deployment, and testing capabilities, which facilitates the reproduction of complex DeFi attack scenarios. The smart contracts used in the experiments are implemented using Solidity. Three representative attack scenarios are constructed, including reentrancy attacks, oracle manipulation attacks, and flash loan-based governance attacks, as shown in Figure ~\ref{fig: Workflow_for_Stablecoin}. For each attack scenario, we first deploy a vulnerable contract implementation and reproduce the corresponding exploit process. Then, the proposed defense mechanism is integrated into the contract and evaluated under the same attack conditions.

For the anomaly detection experiment, the proposed Bi-LSTM model is implemented using PyTorch. Since publicly available datasets containing sufficient labeled stablecoin attack transactions are limited, we construct a simulated transaction dataset based on realistic DeFi attack characteristics. The dataset contains 100,000 transaction samples, including 95,000 normal transactions and 5,000 malicious transactions. Each transaction is represented using multi-dimensional behavioral features, including transaction execution cost, flash loan invocation patterns, oracle price deviation, and contract interaction behaviors.

The dataset is randomly divided into training and testing sets with a ratio of 80\% and 20\%, respectively. The training set is used for model optimization, while the testing set is used for evaluating the generalization capability of the detection model.

\subsection{Evaluation of Smart Contract Defense Mechanisms}

To answer RQ1, we evaluate the effectiveness of the proposed smart contract defense mechanisms against representative attacks targeting stablecoin protocols. Although stablecoin systems differ in their underlying mechanisms, existing attacks generally exploit three fundamental weaknesses, including insecure contract execution logic, unreliable external data sources, and vulnerable economic governance mechanisms. Therefore, we select three representative attack scenarios for evaluation, as shown in Table ~\ref{tab:attack_defense}, including reentrancy attacks, oracle manipulation attacks, and flash loan-based governance attacks.

For each attack scenario, we first deploy a vulnerable smart contract implementation and reproduce the corresponding attack process in the local Ethereum-compatible environment. Subsequently, the proposed defense mechanism is integrated into the vulnerable contract, and the same attack procedure is executed again to evaluate whether the attack can be effectively mitigated. The effectiveness of the defense mechanism is evaluated based on two criteria, including
whether the attacker can successfully complete the malicious transaction and obtain unauthorized benefits, and whether the proposed mechanism can prevent abnormal state transitions while preserving normal contract functionality.

\subsubsection{Reentrancy Attack Evaluation}

Reentrancy attacks represent one of the most common vulnerabilities in Ethereum smart contracts, where an attacker repeatedly invokes a vulnerable external function before the contract updates its internal state. In stablecoin systems, such vulnerabilities may result in unauthorized withdrawal of collateral assets or liquidity drainage from protocol pools.

\textbf{Attack Setup.}
To evaluate the effectiveness of the proposed defense mechanism, we construct a vulnerable stablecoin liquidity pool contract containing a withdrawal function with an insecure execution order. Specifically, the contract transfers assets to the external caller before updating the user's internal balance, creating a reentrancy opportunity.

An attacker contract is deployed to exploit this vulnerability through a fallback function. Once the attacker initiates the withdrawal request, the fallback function is repeatedly triggered, allowing multiple withdrawals within a single transaction before the balance state is updated. The liquidity pool is initialized with sufficient assets to simulate a realistic stablecoin reserve environment.

\textbf{Defense Mechanism.}
To mitigate this vulnerability, we integrate a reentrancy protection mechanism based on execution-state locking. Before executing sensitive external calls, the contract maintains a mutex status variable to prevent recursive invocation of protected functions. Furthermore, the withdrawal logic is modified following the checks-effects-interactions principle. The user's balance state is updated before transferring assets to external addresses, ensuring that subsequent recursive calls cannot reuse the original balance state.

\textbf{Evaluation Result.}
Without the proposed defense mechanism, the attacker contract successfully performs recursive calls and withdraws funds beyond the authorized balance, demonstrating the feasibility of the reentrancy exploit. After applying the reentrancy protection mechanism, the same attack transaction is rejected during execution because the recursive invocation attempt violates the contract state-lock condition. Meanwhile, legitimate withdrawal operations remain unaffected.

The experimental results demonstrate that the proposed defense mechanism effectively prevents reentrancy asset drainage attacks by restricting abnormal recursive execution paths.

\begin{table}[t]
\centering
\caption{Attack Defense Evaluation}
\label{tab:attack_defense}
\begin{tabular}{lll}
\hline
Attack Type & Defense Mechanism & Result \\
\hline

Reentrancy Attack 
& Reentrancy Guard + CEI Principle 
& Prevented \\

Oracle Manipulation 
& TWAP + Multi-source Oracle 
& Mitigated \\

Flash Loan Governance 
& Time-lock Mechanism 
& Prevented \\

\hline
\end{tabular}
\end{table}

\subsubsection{Oracle Manipulation Attack Evaluation}

Oracle manipulation attacks are critical threats to decentralized stablecoin systems because incorrect price information can directly affect collateral valuation, liquidation decisions, and token stability mechanisms. Unlike traditional smart contract vulnerabilities, oracle attacks exploit the dependency between on-chain protocols and external market data sources.

\textbf{Attack Setup.}
To evaluate the effectiveness of the proposed oracle security mechanism, we construct an oracle manipulation scenario based on an AMM liquidity pool. In the vulnerable implementation, the stablecoin protocol directly relies on instantaneous spot prices obtained from a decentralized exchange. An attacker first acquires a large amount of temporary liquidity and performs a series of token swaps within a short period. These operations introduce significant price deviation in the liquidity pool, causing the oracle to return an abnormal asset price. The manipulated price is subsequently utilized by the stablecoin protocol for collateral valuation and liquidation decisions. As a result, attackers can exploit the incorrect price information to obtain unfair liquidation benefits or manipulate protocol states.

\textbf{Defense Mechanism.}
To mitigate oracle manipulation risks, we introduce a TWAP-based oracle mechanism combined with multi-source price aggregation. Instead of directly using instantaneous market prices, the TWAP mechanism calculates asset prices based on historical observations over a predefined time window. Therefore, short-term price fluctuations caused by large transactions cannot immediately influence the reported oracle value. In addition, the multi-source oracle aggregation mechanism collects price information from multiple independent data providers and reduces the impact of a single manipulated price source. A price circuit breaker is further introduced to temporarily suspend sensitive operations when abnormal price deviations exceed predefined thresholds.

\textbf{Evaluation Result.}
In the vulnerable oracle implementation, the attacker can significantly modify the AMM pool price within a single transaction sequence, causing the stablecoin protocol to receive inaccurate price information and execute incorrect financial operations. After integrating the proposed oracle protection mechanism, short-term price manipulation fails to immediately affect the oracle output because the TWAP calculation smooths transient price fluctuations. Meanwhile, abnormal price deviations are detected by the circuit breaker mechanism, preventing potentially harmful liquidation or borrowing operations.

The experimental results demonstrate that the proposed oracle protection strategy effectively improves the resilience of stablecoin systems against short-term price manipulation attacks.

\subsubsection{Flash Loan Governance Attack Evaluation}

Flash loan-based governance attacks have become an emerging threat in decentralized finance systems. Unlike traditional attacks that exploit programming errors, flash loan attacks exploit the atomic execution characteristics of blockchain transactions, allowing attackers to temporarily acquire large amounts of governance power without long-term capital commitment. In stablecoin protocols, such attacks may allow malicious actors to manipulate governance decisions, modify critical protocol parameters, or execute unauthorized administrative operations.

\textbf{Attack Setup.}
To evaluate the effectiveness of the proposed governance protection mechanism, we construct a vulnerable decentralized governance contract combined with a flash loan provider. In the vulnerable implementation, voting power is determined directly based on the current token balance within the same transaction. An attacker can therefore borrow a large amount of governance tokens through a flash loan, temporarily increase their voting weight, submit a malicious proposal, and execute the proposal before the flash loan is repaid. The complete attack process consists of four stages, which include obtaining temporary liquidity through a flash loan, acquiring excessive governance voting power, submitting and approving a malicious proposal, and executing unauthorized governance operations within the same transaction.

\textbf{Defense Mechanism.}
To mitigate flash loan governance attacks, we introduce a time-lock-based governance execution mechanism. Unlike the vulnerable design that allows immediate execution after proposal approval, the proposed mechanism separates proposal approval and execution into two independent phases. After a proposal receives sufficient votes, the execution is delayed for a predefined time interval. During this delay period, the protocol can perform additional security verification, and temporary voting power obtained through flash loans becomes invalid because the borrowed assets must be returned before transaction completion. Therefore, attackers cannot maintain sufficient governance influence during the actual execution phase.

\subsection{Evaluation of AI-based Real-time Attack Detection}

To answer RQ2, we evaluate the effectiveness of the proposed Bi-LSTM-based anomaly detection model in identifying malicious stablecoin transactions. Different from traditional smart contract vulnerability detection methods that analyze source code after deployment, our approach focuses on transaction-level behavioral analysis and aims to detect abnormal activities before transaction confirmation.

\subsubsection{Dataset Construction and Experimental Configuration}

Due to the limited availability of publicly labeled stablecoin attack transactions, we construct a simulated transaction dataset based on representative DeFi attack characteristics and normal transaction behaviors.

The dataset contains 100,000 transaction samples, including 95,000 normal transactions and 5,000 malicious transactions. The malicious samples cover several representative attack patterns, including abnormal contract interactions, flash loan behaviors, oracle price deviations, and irregular transaction execution patterns. Each transaction is represented as a sequential feature vector containing multiple behavioral attributes extracted from transaction execution information. These features include transaction cost, contract invocation patterns, liquidity borrowing behaviors, price deviation information, and temporal interaction characteristics. The dataset is divided into training and testing subsets with an 80:20 ratio for 10 rounds of experiments. During model training, a cost-sensitive binary cross-entropy loss function is adopted to alleviate the impact of class imbalance and improve the detection capability for minority attack samples.

\subsubsection{Detection Performance Evaluation}

We evaluate the detection performance of the proposed model using Accuracy, Precision, Recall, and F1-score. Since attack transactions usually represent only a small proportion of blockchain activities, Recall is considered the most important metric because missing malicious transactions may result in significant financial losses. The experimental results are summarized in Table~\ref{tab:model_performance}.

\begin{table}[htbp]
  \centering
  \caption{Performance Evaluation of the Anomaly Detection Model}
  \label{tab:model_performance}
  \begin{tabular}{cccccc}
    \toprule
    Experiment & Test Set Size & Actual Attacks & Accuracy & Recall & Precision \\
    No.        & (Samples)     & (Samples)      & (\%)     & (\%)   & (\%) \\
    \midrule
    1          & 20000         & 1004           & 96.27    & 97.31  & 57.57 \\
    2          & 20000         & 991            & 97.23    & 97.48  & 64.66 \\
    3          & 20000         & 1015           & 97.91    & 95.76  & 72.11 \\
    4          & 20000         & 1023           & 97.02    & 97.85  & 63.52 \\
    5          & 20000         & 1018           & 95.17    & 98.53  & 51.36 \\
    6          & 20000         & 989            & 96.56    & 98.28  & 59.20 \\
    7          & 20000         & 1058           & 96.70    & 97.83  & 61.90 \\
    8          & 20000         & 972            & 96.74    & 97.53  & 60.15 \\
    9          & 20000         & 1023           & 95.79    & 98.44  & 54.94 \\
    10         & 20000         & 1020           & 96.72    & 97.94  & 61.14 \\
    \midrule
    Average    & 20000         & 1011.3         & 96.61    & 97.70  & 60.66 \\
    \bottomrule
  \end{tabular}
\end{table}

The proposed model achieves an accuracy of 96.61\% and a recall of 97.70\%, demonstrating its effectiveness in identifying malicious transaction behaviors. The high recall indicates that the model can successfully capture most attack transactions, which is particularly important for security-critical blockchain environments where false negatives may lead to irreversible financial losses. However, the precision value is relatively lower compared with recall. This is mainly caused by the highly imbalanced transaction distribution, where certain legitimate high-frequency DeFi operations may exhibit behaviors similar to malicious activities. This trade-off is acceptable because the proposed system prioritizes minimizing missed attacks in real-time protection scenarios.

\subsubsection{Real-time Detection Capability Analysis}

Besides detection accuracy, response latency is another critical factor for blockchain security systems. To evaluate the feasibility of real-time deployment, we measure the inference latency of the trained Bi-LSTM model during transaction-level prediction. The average inference latency ranges from 1.5 ms to 2.8 ms for a single transaction feature sequence. The latency measurement includes feature input processing and model forward inference but excludes network transmission overhead. Compared with the block confirmation interval of Ethereum, which usually requires several seconds, the inference time of the proposed model is sufficiently low to support transaction-level risk monitoring before confirmation. These results demonstrate that the proposed anomaly detection mechanism can be integrated into blockchain nodes or RPC gateways to provide proactive transaction screening.

\section{Discussion}

The experimental results demonstrate that the proposed framework provides an effective and comprehensive security solution for stablecoin smart contracts by combining preventive protection mechanisms with real-time transaction-level anomaly detection. Unlike traditional security approaches that mainly focus on post-deployment vulnerability analysis or isolated attack mitigation, the framework adopts a lifecycle-based defense strategy. The experimental evaluation of reentrancy attacks, oracle manipulation attacks, and flash loan governance attacks shows that different attack surfaces require corresponding protection mechanisms. Specifically, execution-level vulnerabilities can be mitigated through secure coding patterns and runtime protection, while economic and data-related attacks require protocol-level mechanisms such as oracle aggregation and governance delay.

Furthermore, the AI-based detection module complements traditional smart contract security mechanisms by providing proactive protection against transaction-level threats. The Bi-LSTM model achieves a recall of 97.70\%, indicating that it can effectively identify most malicious transaction behaviors before confirmation. This capability is particularly valuable for stablecoin systems because many DeFi attacks exploit transaction ordering, temporary liquidity, or abnormal interaction patterns that cannot be fully detected through static contract analysis alone. However, the current detection model still faces challenges in distinguishing sophisticated malicious transactions from legitimate high-frequency DeFi activities. The relatively lower precision compared with recall indicates that some normal transactions with unusual execution patterns may be incorrectly classified as suspicious behaviors. This limitation mainly originates from the highly imbalanced and dynamic characteristics of blockchain transaction environments. In practical deployment scenarios, the detection module may require adaptive threshold adjustment and continuous model updating to balance security sensitivity and operational efficiency.

From a practical deployment perspective, the proposed framework can be integrated into blockchain nodes, RPC gateways, or DeFi protocol monitoring systems. The low inference latency of the detection model demonstrates its potential for real-time transaction screening. However, further optimization is required to reduce computational overhead and ensure scalability in high-throughput blockchain environments.

\section{Conclusion}

In this paper, we propose a lifecycle-based dynamic security framework for stablecoin smart contracts, integrating preventive smart contract protection mechanisms with real-time transaction-level anomaly detection. The proposed framework addresses multiple attack surfaces, including contract execution vulnerabilities, oracle manipulation, and governance-related threats. Experimental results demonstrate that the proposed defense mechanisms can effectively mitigate representative stablecoin attacks, while the Bi-LSTM-based detection model achieves 96.61\% accuracy and 97.70\% recall with low inference latency, showing its potential for real-time blockchain security monitoring. Although the current evaluation is based on simulated transaction data, this work provides a practical foundation for combining intelligent detection with multi-layer security protection in decentralized financial systems.

\section*{Acknowledgments}
We use AI translation software for grammar improvement and polishing.

\bibliography{main} %
\bibliographystyle{apalike}
\end{document}